\documentclass{PoS}

\usepackage{aas_macros}
\usepackage{lineno}
\usepackage{xspace}
\newcommand{\epem}{$e^+e^-$\xspace}
\newcommand{\gr}{$\gamma$-ray\xspace}

\newcommand{\Grs}{$\gamma$-rays\xspace}

\title{Studying cosmological \gr propagation with the Cherenkov Telescope Array}

\ShortTitle{\gr propagation with CTA}

\author{\speaker{Florian Gat\'e},$^a$ Rafael {Alves Batista},$^{b}$ Jonathan Biteau,$^{c}$ Julien Lefaucheur,$^{d}$ Salvatore Mangano,$^{e}$ Manuel Meyer,$^{f}$ Quentin Piel,$^{a}$ Santiago Pita,$^{h}$ David Sanchez,$^{a}$ and Ievgen Vovk,$^{i}$ for the CTA Consortium\\
\llap{$^a$}Laboratoire d' Annecy-le-Vieux de Physique des Particules (LAPP), Univ. de Savoie, CNRS/IN2P3, Annecy-le-Vieux F-74941, France
\\
\llap{$^b$}Department of Physics - Astrophysics, University of Oxford, DWB, Keble Road, OX1 3RH, Oxford, UK\\
\llap{$^c$}Insitut de Physique Nucl\'eaire d'Orsay, Universit\'e Paris-Sud, Univ. Paris/Saclay, CNRS/IN2P3, 91400 Orsay, France\\
\llap{$^d$}LUTH, Observatoire de Paris, PSL Research University, CNRS, Universit\'e Paris Diderot\\
  5 Place Jules Janssen, 92190 Meudon, France\\
\llap{$^e$}CIEMAT, Avda. Complutense 40, 28040 Madrid, Spain\\
\llap{$^f$}W. W. Hansen Experimental Physics Laboratory,
Kavli  Institute  for  Particle  Astrophysics  and  Cosmology,
Department  of  Physics  and  SLAC  National  Accelerator  Laboratory, Stanford  University,  Stanford,  California  94305,  USA\\
 \llap{$^h$}APC, AstroParticule et Cosmologie, Universit\'{e} Paris Diderot, CNRS/IN2P3, CEA/Irfu,
  Observatoire de Paris, Sorbonne Paris Cit\'{e}\\
  10, rue Alice Domon et L\'{e}onie Duquet, 75205 Paris Cedex 13, France\\
\llap{$^i$}Max Planck Institute for Physics
(Werner-Heisenberg-Institute),
Foehringer Ring 6,
80805 Munich,
Germany\\
E-mail:  \email{florian.gate@lapp.in2p3.fr}, 
\email{biteau@ipno.in2p3.fr},
\email{mameyer@stanford.edu}
}

\abstract{
The measurement of \Grs originating from active galactic nuclei offers the unique opportunity to study the propagation of very-high-energy photons over cosmological distances. Most prominently, \Grs interact with the extragalactic background light (EBL) to produce \epem pairs, imprinting an attenuation signature on \gr spectra. The \epem pairs can also induce electromagnetic cascades whose detectability in \Grs depends on the intergalactic magnetic field (IGMF). Furthermore, physics beyond the Standard Model such as Lorentz invariance violation (LIV) or oscillations between photons and weakly interacting sub-eV particles (WISPs) could affect the propagation of \Grs. The future Cherenkov Telescope Array (CTA), with its unprecedented \gr source sensitivity, as well as enhanced energy and spatial resolution at very high energies, is perfectly suited to study cosmological effects on \gr propagation. Here, we present first results of a study designed to realistically assess the capabilities of CTA to probe the EBL, IGMF, LIV, and WISPs.
}

\FullConference{35th International Cosmic Ray Conference -ICRC2017-\\
		10-20 July, 2017\\
		Bexco, Busan, Korea}

\begin{document}

\section{CTA and the extragalactic sky}

The Cherenkov Telescope Array (CTA, \cite{ksp}) is the next-generation ground-based \gr observatory. With two sites equipped with a total of close to a hundred imaging atmospheric Cherenkov telescopes, the CTA Consortium and guest observers will have access to full-sky coverage, unveiling 
an unprecedented view of the \gr sky from 20\, GeV up to 300\, TeV. CTA observations of extragalactic sources will revolutionize the field of \gr cosmology, which studies
the propagation of the highest-energy photons over cosmological scales.
The CTA Consortium has singled out key science projects (KSP, \cite{ksp}) dedicated to extragalactic science including, but not limited to: an {\bf Extragalactic survey}, covering 25\% of the sky with point source sensitivity down to $\sim 6$\, mCrab above 100\,GeV ($\sim$1000 hours of observations over 10 years) ;  {\bf Transients}, aiming at swift target-of-opportunity observations of, e.g., \gr bursts, which could be detected beyond $z=2$ ; {\bf Active Galactic Nuclei} (AGN), with a long-term monitoring program  of known variable objects ($\sim$1500 hours), a program aiming for a catalog of high-quality spectra  ($\sim$600 hours), and a program aimed at catching flares;
{\bf Clusters of galaxies}, focused in particular on the Perseus cluster, hosting the known \gr emitting AGN IC~310 and NGC~1275 ($\sim$300 hours proposed).

In the following, we illustrate how these dedicated observation programs combined with CTA  instrument response functions (IRFs; e.g. the effective area, energy resolution, angular resolution) will radically transform the field of \gr cosmology.
We use the most up-to-date IRFs (\texttt{prod3}) for both the Northern (La Palma) and Southern (Paranal) CTA sites.

\section{Interaction of \Grs with the extragalactic background light}

The Universe is not fully transparent to \Grs from extragalactic sources. These can annihilate in\epem pairs through interactions with the UV-to-far-infrared photons constituting the extragalactic background light (EBL, \cite{1967PhRv..155.1404G}). The EBL is the second most intense cosmic photon field after the cosmic microwave background (CMB). The amount of \gr absorption can be quantified by the \gr optical depth, $\tau$, which depends on the \gr energy, $E$, and the redshift of the source $z$.

The number of \Grs reaching Earth is reduced by a factor $\exp(-\tau(E,z))$ resulting in an increasingly weaker flux at higher energy and higher redshift. 

\begin{figure}
\centering
\includegraphics[width = 0.85\linewidth]{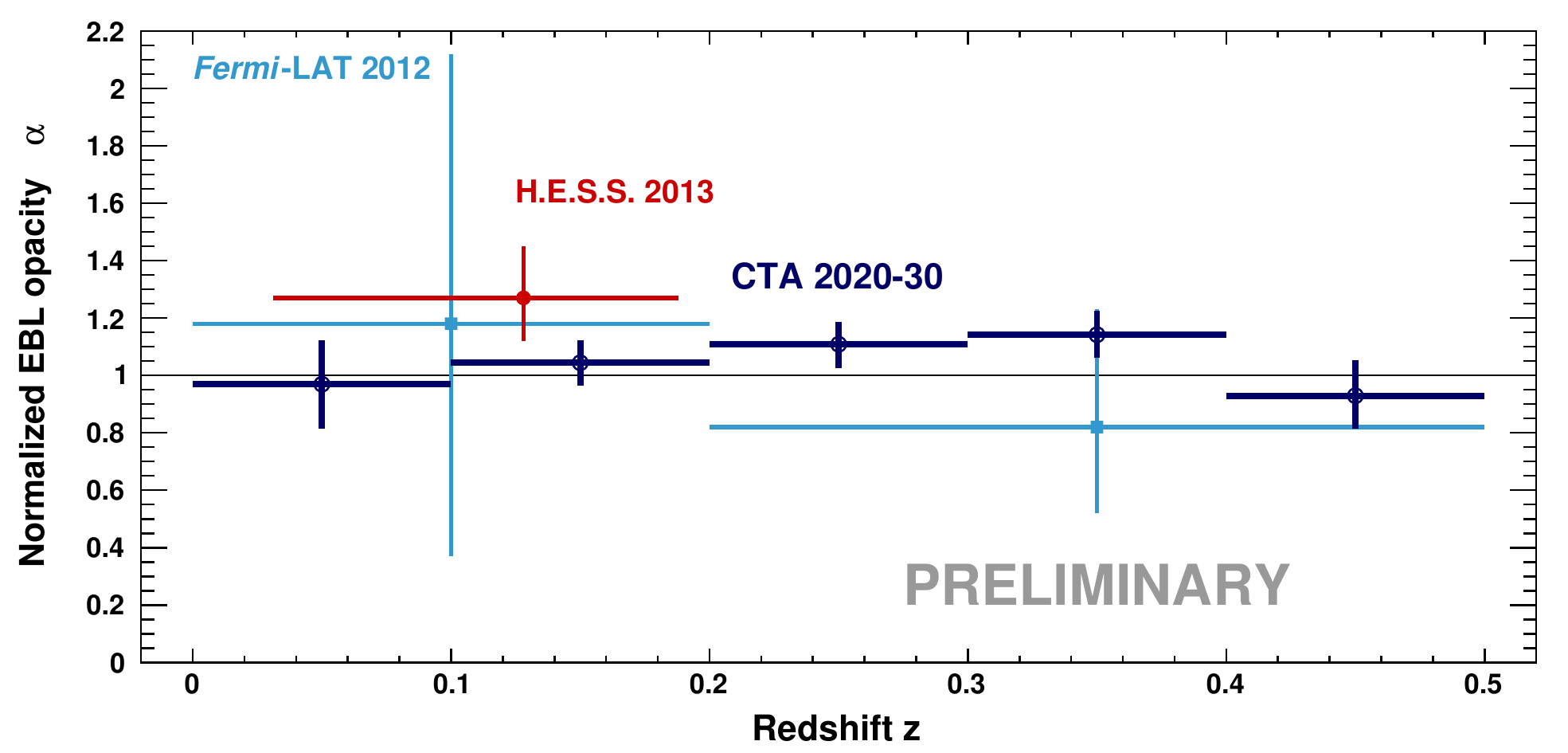}
\caption{Normalization of the EBL vs redshift with respect to a template model, as measured by {\it Fermi}-LAT (light-blue squares) and H.E.S.S. (red circles), and as simulated for CTA (dark blue circles) for observations included in the AGN long-term monitoring program.\label{fig:eblnorm}}
\end{figure}

This distinctive dependence was first extracted in Refs. \cite{2012Sci...338.1190A,2013A&A...550A...4H} by constraining the normalization of template EBL models, illustrated as red and light-blue points in Fig.~\ref{fig:eblnorm}   for the model from Ref. \cite{FR08}. 

We evaluate the capabilities of CTA by simulating observations of 15 long-term monitored sources selected in the AGN KSP~\cite{ksp}, with an exposure of 100~hours per source accumulated over 10 years.
We extract the average GeV-TeV emission of each source from the 3FHL catalog \cite{2017arXiv170200664T}.
Each spectrum is either modeled as a power law or a log-parabola, depending on the significance of the curvature, to which we add an exponential cut-off at $E_{\rm cut}'=1$\,TeV in the frame of the emitting galaxy, that is at $E_{\rm cut}=E_{\rm cut}'/(1+z)$ in the observer's frame. Such a cut-off conservatively reproduces the low flux states observed e.g. by H.E.S.S., MAGIC, and VERITAS. The absorption due to the EBL is in turn modeled according to Ref.~\cite{D11}.
To reconstruct the factor normalizing the amount of absorption due to the EBL, $\alpha$, the intrinsic spectrum of each source is modeled as a power law or log-parabola with an exponential cut-off, using as a first step the intrinsic model selected for the simulation. For sources beyond $z=0.3$, the statistics is not sufficient to probe the intrinsic spectrum in the cut-off regime, so that the cut-off is not included in the model. The intrinsic spectral parameters are left free to vary in the fitting procedure. Absorption is included by multiplying the intrinsic spectra by a factor $\exp(-\alpha\times\tau(E,z))$, where the parameter $\alpha$ is common to multiple sources, grouped by redshift bins of width 0.1.

As shown in Fig.~\ref{fig:eblnorm}, the order-of-magnitude improved sensitivity of CTA, combined with a dedicated AGN observation program, will enable for the first time the probe of the evolution of the EBL opacity at the $\pm10$\,\% level in the redshift range [0 ; 0.5], relying solely on the long-term monitoring program. The study of flaring and transient sources up to and beyond a redshift $z=1$ is left for upcoming developments, together with detailed studies of the spectrum of the EBL.

\section{Deflections of the \epem pairs in the intergalactic magnetic field }
If the \epem~pairs resulting from $\gamma\gamma$ annihilation predominantly cool through inverse-Compton scattering off CMB photons, they will initiate an electromagnetic cascade: the CMB photons are up-scattered to \gr energies and can again pair-produce on the EBL~\cite{protheroe1993}. The spectral, spatial, and temporal signatures of this secondary \gr emission depends on the deflection of the \epem~pairs in the intergalactic magnetic field (IGMF)~\cite{aharonian1994halo,plaga1995} and on the $\gamma$-ray duty cycle of the source~\cite{dermer2011} Therefore, this signal can be used to probe the IGMF strength, $B$, and coherence length, $\lambda$; parameters that are currently only poorly constrained and could trace the possibly primordial origin of cosmic magnetism~\cite{durrer2013}. 

Observations of extreme blazars with hard \gr spectra up to TeV energies are best suited for searches of the cascade emission. \emph{Fermi}-LAT and atmospheric Cherenkov telescope observations of these objects have already been used to limit $B \gtrsim10^{-19}$-$10^{-16}$\,G due to the absence of a spectral bump at GeV energies \cite{finke2015} or a lack of extended $\gamma$-ray emission component around these sources~\cite{veritas2017}. 

\begin{figure}
\centering
\includegraphics[width = 0.54\linewidth]{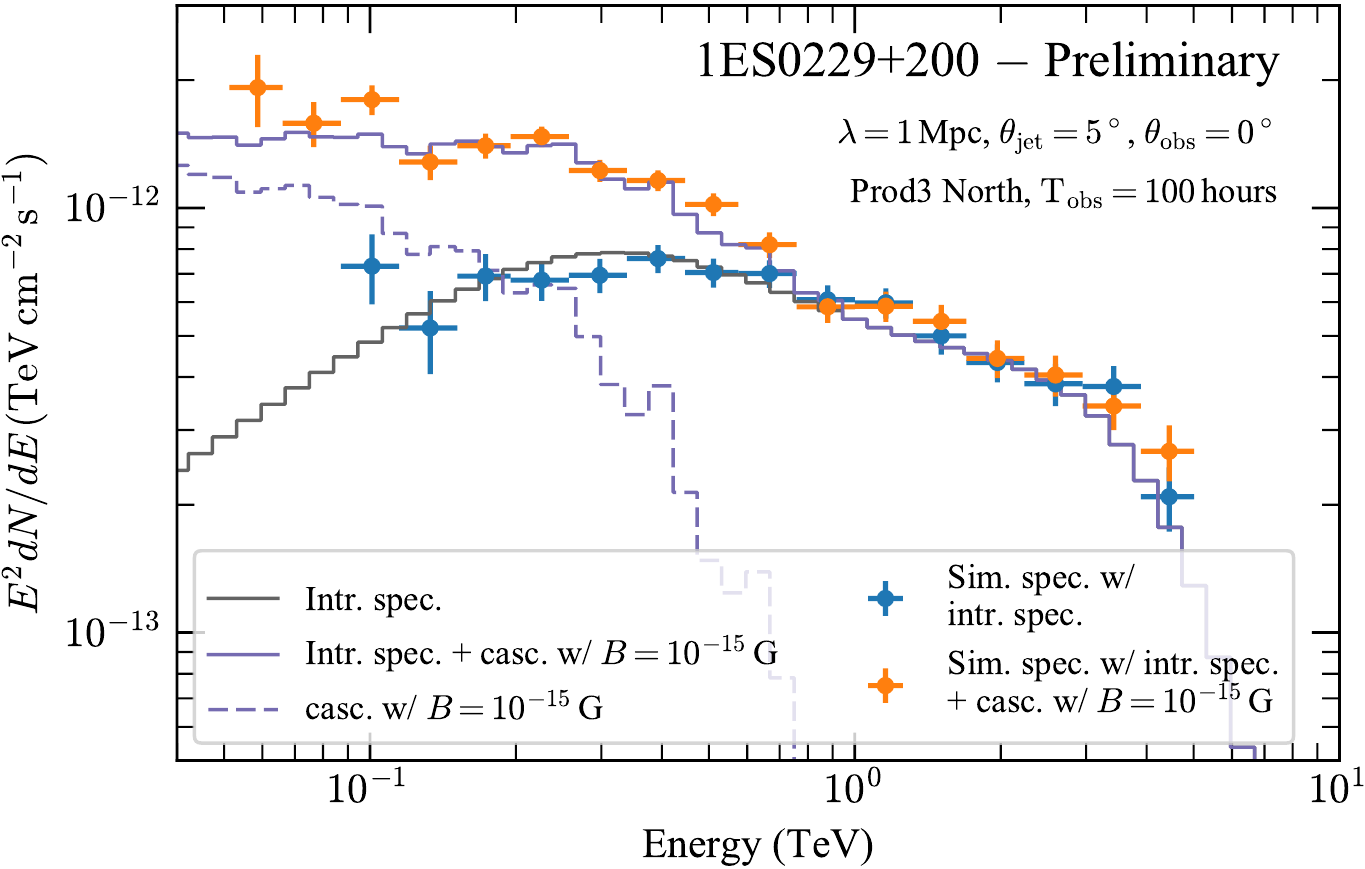}
\includegraphics[width = 0.45\linewidth]{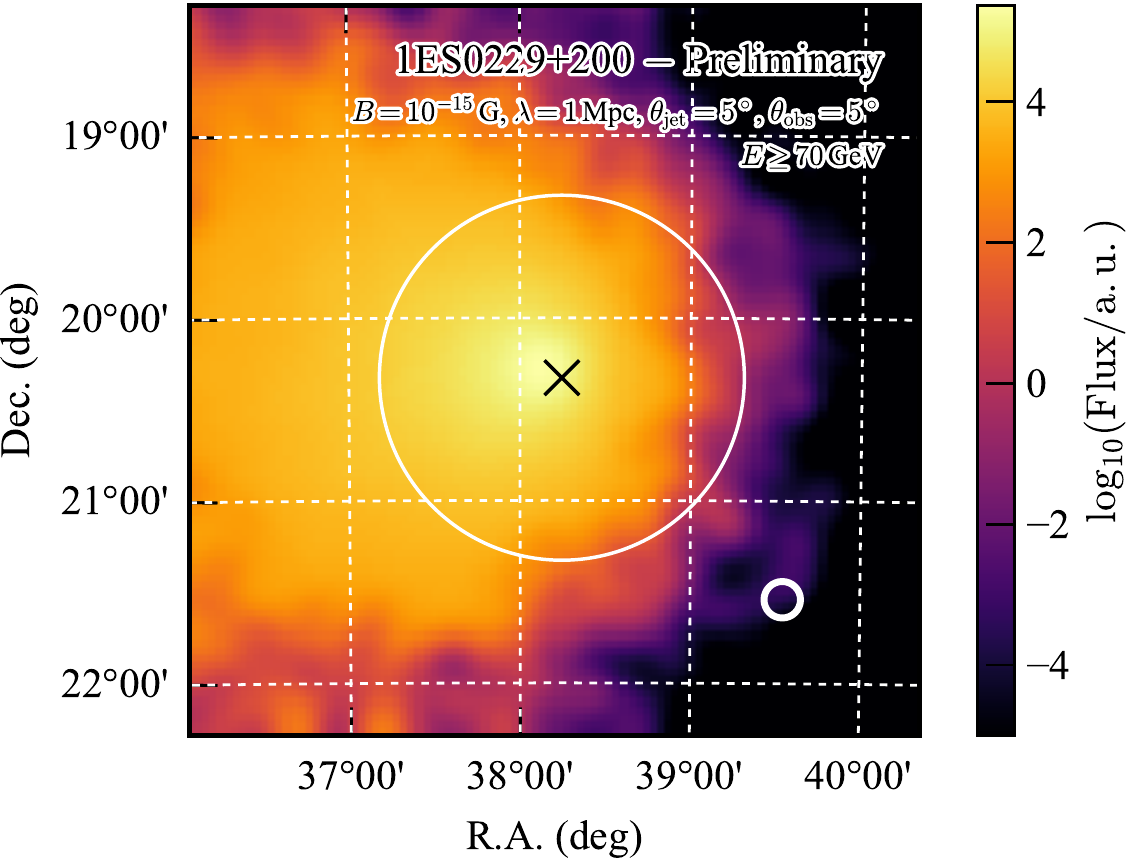}
\caption{\label{fig:igmf}\textit{Left:} Simulated CTA observations of 1ES\,0229+200 with and without secondary cascade emission. \textit{Right:} Simulated sky map for the expected flux in arbitrary units (a.u.) for the primary point-source and secondary extended cascade emission as obtained from the CRPropa simulation. The cross marks the point-source position and the ring around it the integration radius. The small ring in the lower right indicates the radius used for smoothing.}
\end{figure}

With its improved sensitivity, energy range, and spatial resolution, CTA will be perfectly suited to search for cascade emission~\cite{meyer2016cta}. In the left panel of Fig.~\ref{fig:igmf}, we show 
simulated CTA observations of the extreme blazar 1ES\,0229+200, using the CRPropa 3D Monte-Carlo code to simulate the cascade~\cite{crpopa2016}.
As an example, we assume $B = 10^{-15}$\,G and $\lambda = 1\,$Mpc and that the source has been active long enough such that all cascade photons have reached Earth. The intrinsic spectrum is described by a power law with index $\Gamma = 1.2$ and an exponential cut-off at 5\,TeV and the EBL absorption is taken from Ref.~\cite{FR08}. A jet opening angle of $\theta_\mathrm{jet} = 5^\circ$ is assumed while the observation angle is set to $\theta_\mathrm{obs} =0^\circ$. Signal and background events are integrated within a region with a radius of $1^\circ$. The excess of secondary \Grs is clearly visible below $\sim 500$\,GeV in comparison to the spectrum simulated without the cascade contribution. 
The spatial profile of the CRPropa simulation output is shown in the right panel of Fig.~\ref{fig:igmf} for energies above 70\,GeV. The simulation output has been smoothed with a 2D Gaussian with a width of $0.1^\circ$, which roughly corresponds to the envisioned CTA point spread function above 100\,GeV. Assuming now $\theta_\mathrm{jet} = \theta_\mathrm{obs} = 5^\circ$ gives rise to the asymmetric shape of the emission (in accordance with e.g. Ref.~\cite{neronov2010jets}). 

Dedicated analysis tools are under development to simultaneously search for the spectral and spatial cascade signals, as predicted from these simulations. 
A comprehensive scan of the IGMF parameter space, allowing also for a turbulent or helical IGMF, together with a likelihood stacking analysis of numerous extreme blazars that will be observed with CTA, promise the best sensitivity for a detection of an IGMF signature so far. 

\section{Coupling of \Grs to WISPs}
Measuring \Grs from distant galaxies also probes the interaction between photons and weakly interacting sub-eV particles (WISPs, e.g. axions). 
The photon-WISP interaction can leave unique features in the spectra of AGN, as photons oscillate into such particles in the presence of electro-magnetic fields~\cite{raffelt1988}. Such particles arise in various extension of the standard model and are well-motivated particle candidates for cold dark matter (see e.g. Ref.~\cite{arias2012}).

Besides extending the \gr horizon, one observable feature of photon-WISP oscillations are spectral irregularities that should be imprinted on spectra around a critical energy, $E_\mathrm{crit}
\sim 2.5\,\mathrm{GeV}|m_\mathrm{neV}^2 - \omega_\mathrm{pl,\,neV}^2| / g_{11} B_{\mu\mathrm{G}}$
 (e.g. Ref.~\cite{hooper2007:alps}), where $m_\mathrm{neV} = m_a / \mathrm{neV}$ is the WISP mass $m_a$ in neV, 
 $\omega_\mathrm{pl,\,neV}$ is the plasma frequency of the medium in neV, $g_{11} = g_{a\gamma} / 10^{-11}\,\mathrm{GeV}^{-1}$ is the photon-WISP coupling, and $B_{\mu\mathrm{G}}$ is the external magnetic field in $\mu\mathrm{G}$. 
The absence of irregularities in X-ray and \gr spectra of AGN in centers of galaxy clusters have already provided strong limits on $g_{a\gamma}$ for masses below $m_a \lesssim 20\,$neV 
\cite{TheFermi-LAT:2016zue,Wouters:2013hua}. 
With its large collection area and energy coverage, 
CTA is perfectly suited to extend WISP searches to regions where WISPs could constitute the entirety of dark matter. 

\begin{figure}
\centering
\begin{minipage}{0.49\linewidth}
\centering
\includegraphics[width = .99\linewidth]{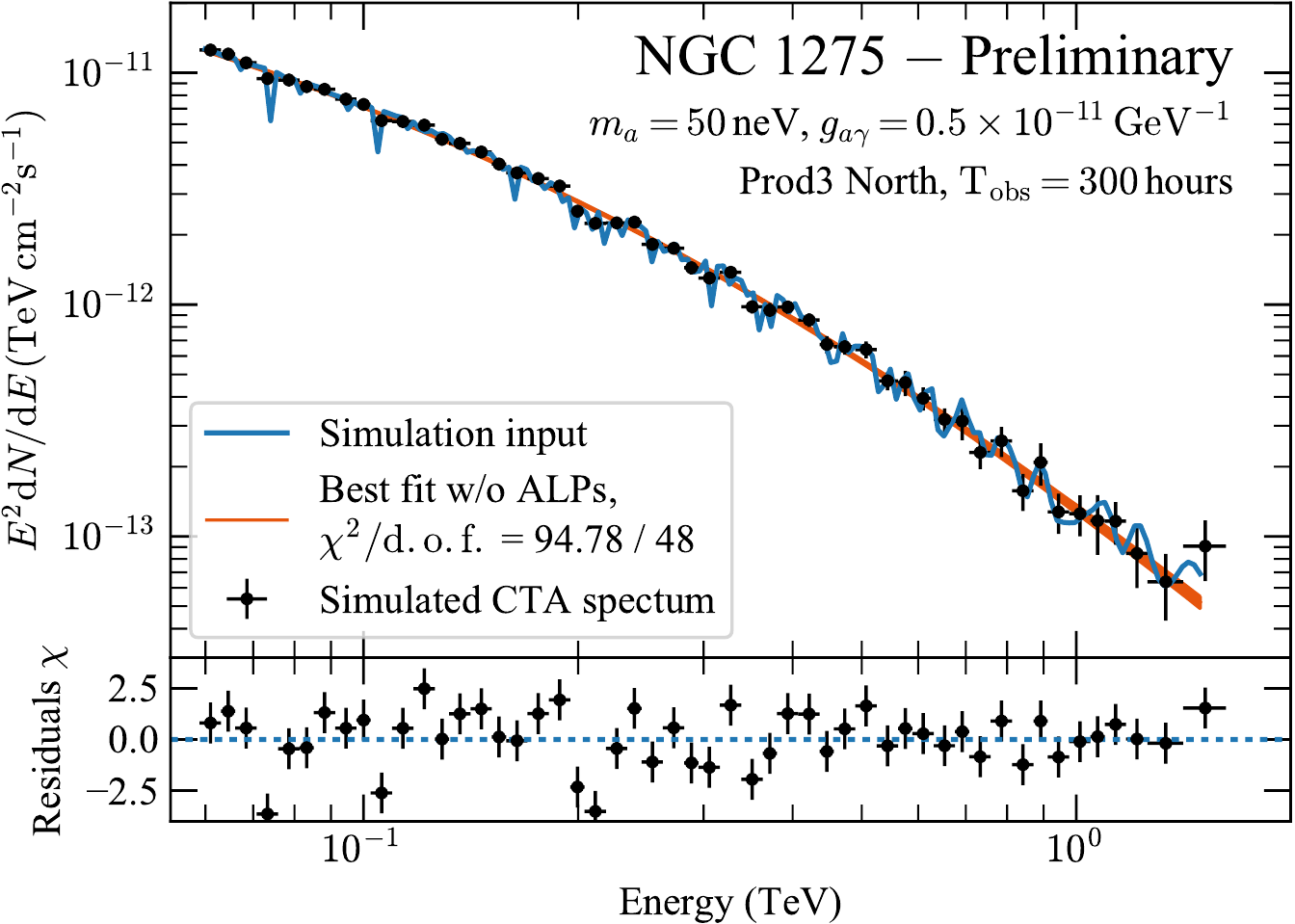}
\end{minipage}
\quad
\begin{minipage}{0.29\linewidth}
\caption{\label{fig:ngc1275alps}
Simulated observation of NGC\,1275 including the effect of photon-WISP oscillations in one random realization
of the turbulent magnetic field of the Perseus cluster 
and the coherent magnetic field of the Milky Way.}
\end{minipage}
\end{figure}

Figure~\ref{fig:ngc1275alps} shows a simulated observation of NGC\,1275, the central galaxy of the Perseus cluster, including photon-WISP oscillations. We adopt the same assumptions for the cluster and Milky Way magnetic fields as in Ref.~\cite{TheFermi-LAT:2016zue} and use a broken power law to model the intrinsic spectrum of the AGN. The spectral parameters are chosen such that the spectrum connects the spectra observed with \emph{Fermi}-LAT~\cite{TheFermi-LAT:2016zue} and MAGIC~\cite{Ahnen:2016qkt}. 
The MAGIC observations revealed a steep power law with index $\Gamma=3.8$ between 90\,GeV and 1.2\,TeV.
We assume a single possible realization of the turbulent magnetic field and WISP parameters $m_a = 50\,$neV and $g_{a\gamma} = 5\times 10^{-12}\,\mathrm{GeV}$. These parameters are currently not probed by any experiment and lie within the region where WISPs could constitute the entire dark matter content of the Universe.
The simulated spectrum including spectral irregularities is shown as a blue solid line in Fig.~\ref{fig:ngc1275alps}. 
Using the latest IRFs and assuming an observation time of 300\,hours, as currently foreseen in the galaxy cluster KSP~\cite{ksp}, 
results in the simulated black data points. 
A fit with a smooth log parabola (orange band) reveals distinctive residuals due to the photon-WISP oscillations and a large $\chi^2 /\mathrm{d.o.f.} = 94.8 / 48$, which corresponds to a $p$-value of $7\times10^{-5}$. This preliminary result demonstrates that CTA will be able to probe new regions of the WISP parameter space. Following e.g. Ref.~\cite{TheFermi-LAT:2016zue}, dedicated analysis techniques will be developed in the future in order to identify the full ($m_a$, $g_{a\gamma}$) parameter space that can be probed with CTA.

\section{Probing physics up to the Planck scale and above}

Finally, studying \Grs at the highest energies enables the search for Lorentz invariance violation (LIV). While quantum-gravity phenomenology is still an exploratory field, a leading-order modification of the photon (and lepton) dispersion relation could become important close to a so-called quantum gravity energy scale, $E_{\rm QG}$, around e.g. the Planck scale, $E_{\rm Planck} = \sqrt{\hbar c^5/G} = 1.22 \times 10^{28}$~eV \cite{2013LRR....16....5A}. While such an effect has often been searched for in the time domain (e.g. with GRBs \cite{2013PhRvD..87l2001V}, or AGNs and pulsars \cite{LIVproc}), which assumes that the Hamiltonian relation $v = \partial E / \partial p$ is still valid close to $E_{\rm QG}$, a complementary channel has been identified in the modification of the pair-creation threshold, which in turns assumes conservation of the four-vector momentum in a preferred frame \cite{2008PhRvD..78l4010J,2014JCAP...06..005F,2016A&A...585A..25T}.

For subluminal effects and assuming that only the photons are affected, the threshold for pair-creation is modified as $\epsilon_{thr} > m_e^2 c^4 / E \times[1+ (E / E_{\gamma, \rm LIV})^{n+2}] $, where $\epsilon_{thr}$ is the threshold EBL photon energy, $E$ the \gr energy, $n$ the order of the modification ($n=1,2$ for a linear/quadratic modification), and $E_{\gamma, \rm LIV} = [4m_e^2c^4E_{\rm QG}^{n}]^{1/(n+2)}$. Such a modification would thus result in an increase of the EBL photon threshold energy, hence a decrease in the number of target photons with which \Grs can interact, yielding an anomalous transparency of the Universe to \Grs above a few tens of TeV. Limits at the Planck scale level have already been obtained for a linear modification, and at the level of a few $10^{20}$~eV for a quadratic modification (e.g. Ref.~\cite{2015ApJ...812...60B}). 

\begin{figure}
\centering
\includegraphics[width = .49\linewidth]{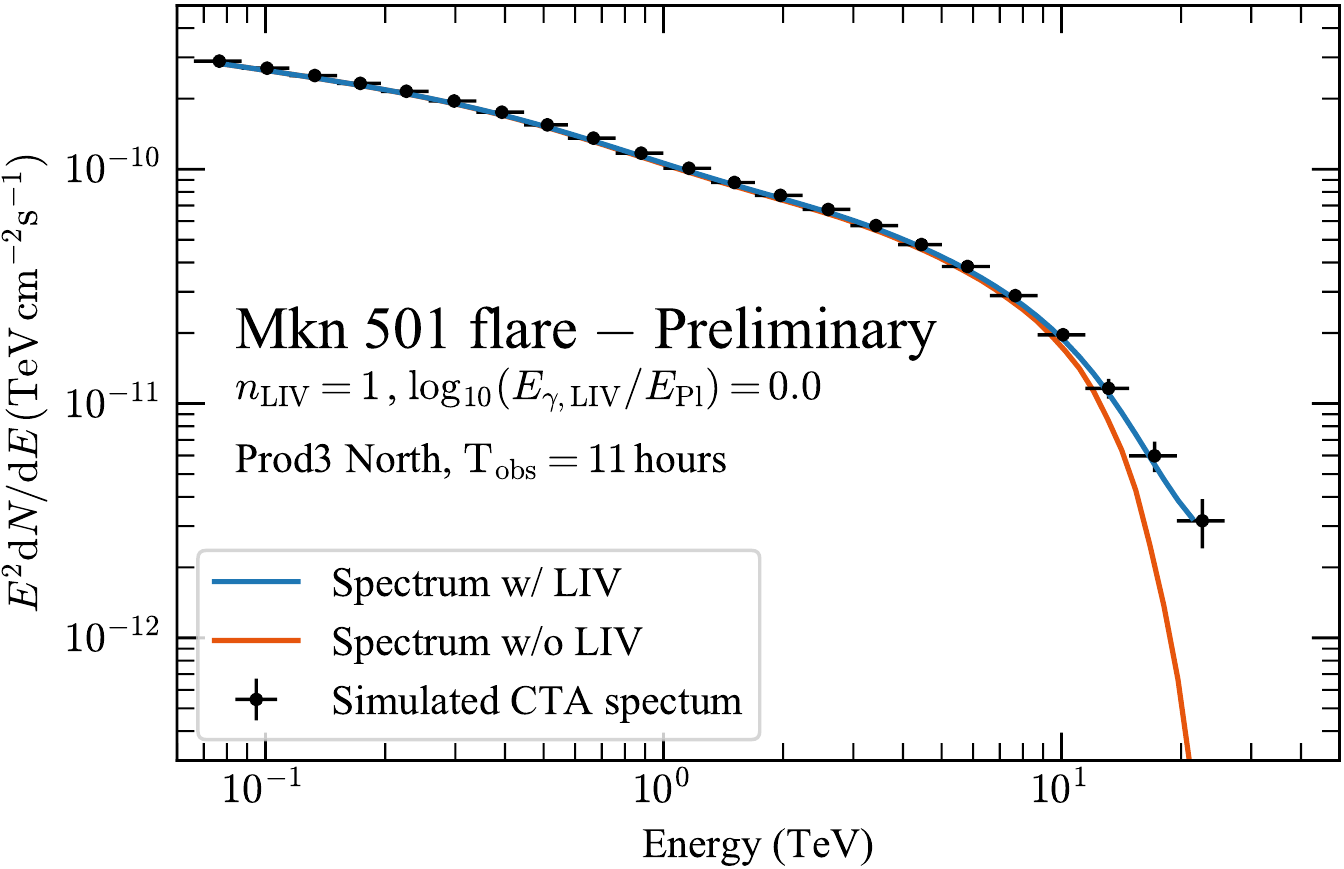}
\hfill
\includegraphics[width = .49\linewidth]{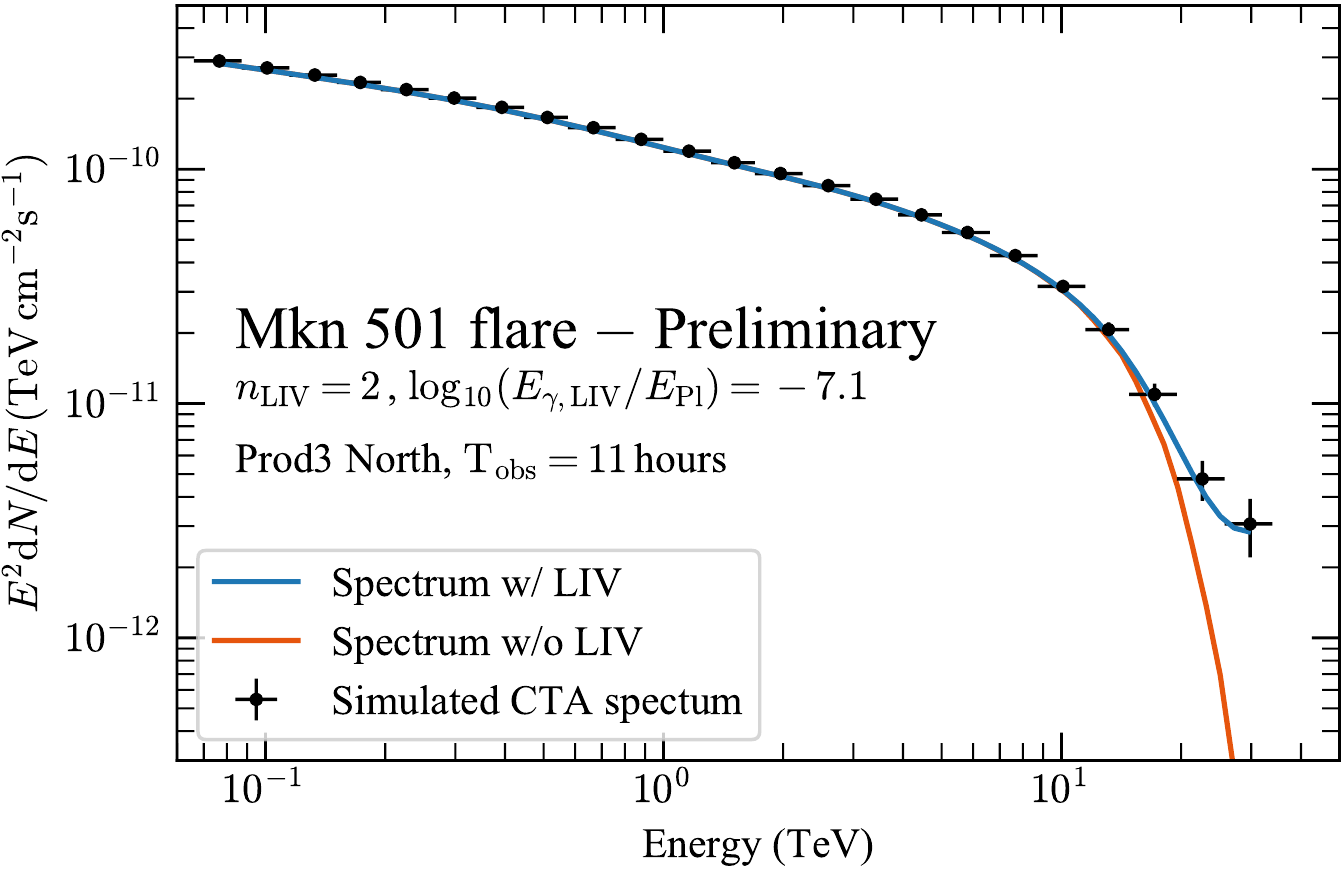}
\caption{
Simulated observation of Mrk\,501 during a flaring state including a linear (left) and quadratic (right) LIV modification of the pair creation threshold.\label{fig:mkn501liv}}
\end{figure}

Figure~\ref{fig:mkn501liv} illustrates the potential of CTA to probe linear (left) and quadratic (right) modifications. We simulate the spectrum of the nearby blazar Mrk~501 ($z=0.034$) observed by HEGRA in 1997~\cite{2001AA...366...62A}. Correcting for EBL absorption as in Ref.~\cite{2015ApJ...812...60B} results in an intrinsic spectrum well described by a power-law of index $\Gamma=2.24$ up to 21~TeV. We conservatively add an exponential cut-off at 20 TeV to avoid overoptimistic predictions at the highest energies. Order of magnitude deviations from classical absorption can be observed in the simulated spectrum between 20 and 30 TeV, assuming a modification at the Planck scale for the linear correction and at $E_{\rm QG}=10^{21}$~eV for the quadratic one. Following e.g. Ref.~\cite{2015ApJ...812...60B}, a scan as a function of $E_{\rm QG}$ will be developed to determine the constraints that can be expected from CTA observations of AGN above tens of TeV. 

\section{Conclusion}
We have presented a preliminary study of the CTA sensitivity to signatures imprinted on \gr spectra due to a variety of effects that might affect the propagation of \Grs over cosmological distances.
The results make use of the latest CTA IRFs and state-of-the-art numerical calculations to evaluate the expected signals in the different tested scenarios.
Albeit preliminary, our findings already demonstrate the capability of CTA to probe various science cases ranging from cosmology (EBL and IGMF) to 
fundamental physics (WISP dark matter and LIV) with unprecedented sensitivity. 
A systematic study that will address the full potential of CTA and the development of dedicated analysis techniques is currently in preparation within the CTA Consortium.

\acknowledgments
This work was conducted in the context of the CTA Physics Working Group.
We gratefully acknowledge financial support from the agencies and organizations listed here: \href{http://www.cta-observatory.org/consortium_acknowledgments/}{http://www.cta-observatory.org/consortium\_acknowledgments/}. MM is a Feodor-Lynen Fellow and acknowledges support of the Alexander von Humboldt Foundation.

\bibliographystyle{JHEP}
\bibliography{cta_gpropa}

\end{document}